\documentclass[amstex,amsfonts,aps,prb,overcite,twocolumn,amsmath]{revtex4}
\usepackage{graphicx}

\begin{document}
\title{\textbf{Reconsideration of Second Harmonic Generation from neat Air/Water Interface:
Broken of Kleinman Symmetry from Dipolar Contribution}}
\author{\textsf{Wen-kai Zhang}\footnote[2]{Also Graduate School of the Chinese
Academy of Sciences}}
\author{\textsf{De-sheng Zheng\footnotemark[2]}}
\author{\textsf{Yan-yan Xu\footnotemark[2]}}
\author{\textsf{Hong-tao Bian\footnotemark[2]}}
\author{\textsf{Yuan Guo}}
\author{\textsf{Hong-fei Wang}\footnote[1]{Author to whom correspondence should be
addressed. E-mail: hongfei@mrdlab.icas.ac.cn. Tel. 86-10-62555347,
Fax 86-10-62563167.}}\affiliation{State Key Laboratory of
Molecular Reaction Dynamics,
\\Institute of Chemistry, the Chinese Academy of Sciences,
Beijing, China, 100080}

\date{\today}

\begin{abstract}
It has been generally accepted that there are significant
quadrupolar and bulk contributions to the second harmonic
generation (SHG) reflected from the neat air/water interface, as
well as common liquid interfaces. Because there has been no
general methodology to determine the quadrupolar and bulk
contributions to the SHG signal from a liquid interface, this
conclusion was reached based on the following two experimental
phenomena. Namely, the broken of the macroscopic Kleinman
symmetry, and the significant temperature dependence of the SHG
signal from the neat air/water interface. However, because sum
frequency generation vibrational spectroscopy (SFG-VS) measurement
of the neat air/water interface observed no apparent temperature
dependence, the temperature dependence in the SHG measurement has
been reexamined and proven to be an experimental artifact. Here we
present a complete microscopic analysis of the susceptibility
tensors of the air/water interface, and show that dipolar
contribution alone can be used to address the issue of broken of
the macroscopic Kleinman symmetry at the neat air/water interface.
Using this analysis, the orientation of the water molecules at the
interface can be obtained, and it is consistent with the
measurement from SFG-VS. Therefore, the key rationales to conclude
significantly quadrupolar and bulk contributions to the SHG signal
of the neat air/water interface can no longer be considered as
valid as before. This new understanding of the air/water interface
can shed light on our understanding of the nonlinear optical
responses from other molecular interfaces as well.
\end{abstract} 

\maketitle

\section{Introduction}

Since Shen and his co-workers established that the second harmonic
generation (SHG) from an interface can be dominated with the
electric dipolar contribution of the interfacial molecular layers,
SHG has been widely used as the spectroscopic probe for molecular
orientation, structure, spectroscopy and dynamics of liquid/vapor
interfaces as well as other interfaces, because SHG is
electric-dipole forbidden in the centrosymmetric bulk media.
\cite{Shen1983PRA28p1883,Richmond1985Review,Eisenthal1988JPC92P5074,
ShenARPC1989,Eisenthal1989CPL157p101,EisenthalAccChemRes1993,Eisenthal1993CPL202p513,
Richmond1994JPC98p9688,Girault1995JCSFT91p1763,Eisenthal1996ChemRev,
Girault1997JCSFT93p3833,Frey2001MP99p677} In the early days of the
development of SHG for interface studies, it has been fully
realized that, in applying SHG to real problems, it is important
to know how to assess the relative magnitudes of the interface
(local) contribution and the still possibly significant bulk
electric-quadrupole and magnetic-dipole (nonlocal) contributions
to the SHG signal.
\cite{ShenARPC1989,ShenGuyotSionnestPRB1986,ShenGuyotSionnestPRB1987,
ShenGuyotSionnestPRB1988} Even though there were exchanges of
debates on whether SHG can be an effective probe for isotropic
liquid interfaces, i.e. whether the bulk contribution is
negligible,\cite{AndrewsPRA1988,ShenPRA1990,AndrewsPRA1990,HeinzPRA1990,
AndrewsJModOpt1993} it has been generally accepted that it is
impossible to separate the bulk contribution from the total SHG
signal.\cite{ShenGuyotSionnestPRB1988,HeinzPRA1990,HeinzBook} Over
the past two decades, theoretical treatment has shown that there
is no general solution to this problem, and it is often not known
\textit{a priori} in interface SHG and sum frequency generation
vibrational spectroscopy (SFG-VS) studies whether in an interface
system interfacial contribution is dominant over that of the bulk,
or not. \cite{ShenAppliedPhysics1999,ShenheldPRB2002} Recently,
Morita proposed a formulation towards computation of quadrupolar
contribution in SFG-VS, which can also be used for
SHG.\cite{MoritaCPL2005SFG} However, no computational result has
been reported so far.

On the other hand, a number of experimental studies have indicated
that `...in the case of insulators and liquids with
$\epsilon_{2}\sim 3$, the bulk contribution may still be
appreciable.', \cite{ShenGuyotSionnestPRB1986} and `For small
adsorbed molecules or symmetric molecules, or molecules with no
delocalized electrons, the electric-quadrupole nonlinearity could
be dominant if both $\omega$ and $2\omega$ are away from resonance
with electric-dipole-allowed
transitions.'\cite{ShenGuyotSionnestPRB1987} Thus, it generally
nullified the possibility for using non-resonant SHG to obtain
detailed molecular information for pure liquid interfaces, and the
SHG study on the neat air/water interface is one of the benchmark
examples. This is also why the majority of literatures on SHG
applications are mainly on near resonant chromophores at various
interfaces.

The first neat liquid interface studied with SHG experiment is the
neat air/water interface by Goh \emph{et
al.}.\cite{Eisenthal1988JPC92P5074} In this and the following
works, two basic phenomena of the neat air/water interface were
observed. One is the significant temperature dependence of the SHG
signal (the SHG intensity dropped more than 50$\%$ when
temperature was changed from 283K to 353K), and the other is the
broken of the Kleinman symmetry, i.e. $\chi_{xzx}\approx
2.3\chi_{zxx}$ at
293K.\cite{Eisenthal1988JPC92P5074,Eisenthal1989CPL157p101} Since
it was generally believed that the Kleinman symmetry rule for
dipolar susceptibility states that in the static limit, i.e. far
from electronic resonance, there is
$\chi_{xzx}=\chi_{zxx}$,\cite{Kleinman_1962PR126p1977} the broken
of Kleinman symmetry for the neat air/water interface was believed
to indicate a significant quadrupole (bulk)
contribution.\cite{Eisenthal1988JPC92P5074,Eisenthal1989CPL157p101}
In addition to the broken of Kleinman symmetry, the strong
temperature dependence was also believed to indicate significant
quadrupolar contribution. The temperature dependence was further
used to separate the dipolar (interface or local) contributions
and the quadrupolar (bulk or nonlocal) contributions, because the
dipolar part should strongly depend on the absolute orientation of
the water molecules, whereas the quadrupolar contribution should
only be weakly dependent on
such.\cite{Eisenthal1988JPC92P5074,Eisenthal1989CPL157p101} Using
a simple model, Goh \textit{et al.} further estimated that the
energy that causes a net absolute orientation of water molecules
at the interface to be about $\frac{1}{2}kT$ at room
temperature.\cite{Eisenthal1989CPL157p101}

However, the temperature dependence result in SHG measurement was
questioned by Du \textit{et al.} with the first SFG-VS
investigation of the neat air/water interface, where the
vibrational spectra of the interfacial free OH group was directly
measured.\cite{ShenDuQ1993PRL70p2313} Temperature dependence of
the SFG spectrum of the air/water interface showed no significant
change over the same temperature range from 283K to
353K.\cite{ShenDuQ1993PRL70p2313} As reported, Du \textit{et al.}
also repeated the SHG temperature dependent experiment of the neat
air/water interface from 283K to 318K, and they observed no change
of $\chi_{xzx}$ within the experimental uncertainty of $\pm 10\%$,
while a $50\%$ decrease was expected from the Goh \textit{et
al.}'s
results.\cite{Eisenthal1988JPC92P5074,Eisenthal1989CPL157p101,ShenDuQ1993PRL70p2313}

On the request by Professor Y. R. Shen, Professor K. B.
Eisenthal's group repeated the SHG temperature experiment of the
neat air/water interface in January of 1993, and found that in the
same temperature range, $\chi_{xzx}$ decreases only by less than
$10\%$ and $\chi_{zxx}$ decreases by about $10\%$, as reported in
Du \textit{et al.}.\cite{ShenDuQ1993PRL70p2313} One of us
(Hong-fei Wang) was then the student in the Eisenthal group who
was responsible for repeating the SHG temperature dependence
measurement. It was found that the strong SHG temperature
dependence reported
previously\cite{Eisenthal1988JPC92P5074,Eisenthal1989CPL157p101}
was an experimental artifact due to the uncontrolled condensation
of water vapor on the internal surfaces of the two quartz windows
of the enclosed temperature controlled cell made of brass. The
condensation became visible when the internal temperature of the
cell is higher than the room temperature controlled at
$22.0\pm1.0^{\circ}$C, and more condensation was formed when the
temperature difference became larger before it reached a steady
state. Thus, the SHG signal drop became more pronounced with the
increase of the degree of condensation until it levelled off
around 310K. Heating of the external surface of the temperature
cell with slightly higher external temperature than the internal
temperature can prevent condensation on the internal surface of
the quartz windows. After getting rid of the condensation, no SHG
signal temperature dependence was observed beyond the experimental
uncertainties, which were $7\%$ for the $\chi_{xzx}$ and $10\%$
for the $\chi_{zxx}$, which is significantly
smaller.\cite{Notebook} Thus, the strong SHG temperature
dependence reported
previously\cite{Eisenthal1988JPC92P5074,Eisenthal1989CPL157p101}
was an experimental artifact, and was not consistent with the
SFG-VS measurement.\cite{ShenDuQ1993PRL70p2313} Recently, with the
knowledge that Goh \textit{et al.}'s SHG temperature dependence
results were not repeatable, Fordyce \textit{et al.} reported a
new measurement that there was a $20\%$ of the SHG signal
temperature drop of the neat air/water interface from 293K to
343K.\cite{Frey2001MP99p677} Careful examination of this report
makes us believe that without effective condensation control for
the temperature cell windows, this work might have also been
subjected to the same experimental artifact as that of Goh
\textit{et al.}. Nevertheless, all the SHG experimental results
confirmed the validity of the broken of Kleinman symmetry for the
neat air/water
interface.\cite{Eisenthal1988JPC92P5074,Eisenthal1989CPL157p101,
Frey2001MP99p677,Girault1995JCSFT91p1763}

Because the magnitude of the temperature dependence of the SHG
signal was believed to indicate the magnitude of the dipolar
(local) contribution to the SHG
signal,\cite{Eisenthal1988JPC92P5074,Eisenthal1989CPL157p101,Frey2001MP99p677}
no apparent temperature dependence could lead to the conclusion
that the apparent dipolar contribution to the SHG signal from the
neat air/water interface is negligible. In addition, the broken of
the Kleinman symmetry was also an indication of significant
quadrupolar (nonlocal) contribution to the SHG signal from the
neat air/water interface. Therefore, it seems reasonable to
conclude that all the SHG signal from the neat air/water interface
was quadrupolar contribution in nature$!$ Indeed, Bloembergen
\textit{et al.} did conclude as early as 1968 that the `SHG
reflection from media with inversion symmetry is described rather
well by the quadrupole-type nonlinear properties calculated for
the homogeneous bulk material with a abrupt discontinuity at the
boundary.'\cite{BloembergenPR1968} However, Bloembergen \textit{et
al.}'s discussion was on experimental results and theoretical
analysis of metal and alkali halides surfaces, which possess
relatively high density of valence electrons, but not on surfaces
of normal liquid. Such a strong statement on metal and alkali
halides surfaces by Bloembergen \textit{et al.} was firstly
questioned by Brown \textit{et al}'s SHG experiment on adsorbed
surface layer on silver in 1969,\cite{BrownPhysRev1969} but only
fundamentally modified more than a decade later, after Y. R.
Shen's group demonstrated systematically in the early 1980's that
surface SHG can be treated as radiation from a nonlinear
polarization sheet induced by an incoming wave at the surface,
besides comparable quadrupole (bulk)
contributions.\cite{ARPC1989Cited12,ARPC1989Cited131,ARPC1989Cited132,ARPC1989Cited14,ShenARPC1989}
This development undoubtedly assured the surface monolayer
sensitivity of surface SHG for all kinds of
surfaces.\cite{ARPC1989Cited12,ARPC1989Cited131,ARPC1989Cited132,ARPC1989Cited14,ShenARPC1989}
These works by Y. R. Shen and his colleagues heralded the
beginning of systematic interface studies with nonlinear optical
techniques,\cite{ShenARPC1989} including studies of the longtime
nail-biting liquid interfaces.\cite{EisenthalAccChemRes1993} These
twists in the historical development of surface SHG demonstrated
the complexity and difficulty on the evaluation of relative bulk
and surface contributions to surface SHG signal.

Now, if the above conjecture on the pure quadrupolar (bulk)
contribution to SHG from neat air/water interface could be held
true, the observed SHG signal could have no surface specificity
and should not have been used to study the molecular orientation
and structure of the air/water interface. And this can also be the
same for other liquid interfaces. The consequence is that the
whole effort of using SHG for non-resonant liquid interface
studies should have been nullified. Indeed, in the past decade,
using SHG to study neat air/liquid and liquid/liquid interfaces
has been
rare.\cite{Girault1995JCSFT91p1763,Girault1997JCSFT93p3833,Frey2001MP99p677}
This situation would persist unless the widely held
assumption,\cite{Eisenthal1988JPC92P5074,Eisenthal1989CPL157p101,
Frey2001MP99p677,Girault1995JCSFT91p1763} that the broken of the
Kleinman symmetry is an indication of significant quadrupolar
(nonlocal) contribution to the SHG signal from an interface, is
generally not true. Examination of the history of the Kleinman
symmetry of the nonlinear microscopic polarizability and
macroscopic susceptibility tensors indeed raised more questions on
the general validity of this symmetry relationship which first
discussed by Kleinman in
1962.\cite{Kleinman_1962PR126p1977,BoydNonlinearOpticsBook}

The validity of the Kleinman symmetry of the second order
nonlinear susceptibility tensors was thoroughly discussed by
Franken and Ward as early as in 1963,\cite{Franken1963RMP35p23}
and most recently by Dailey \textit{et
al.}\cite{Simpson2004CPL390p8} As pointed out by Franken and Ward,
Kleinman suggested that for a material transparent (far from
resonance) to the fundamental and the SH frequencies, an
additional symmetry condition of $\chi_{ijk}=\chi_{jik}$ exists,
where the first index is for the SH frequency, and the second and
third indexes are for the fundamental
frequency.\cite{Franken1963RMP35p23} To translate this for water,
$\chi_{xzx}=\chi_{zxx}$ should hold be true if the Kleinman
symmetry is valid when both fundamental and the SH frequencies are
far from resonance with the water molecule transition frequencies.
It was well recognized by Franken and Ward that the presence of
dispersion can destroy the validity of the Kleinman symmetry, and
using a quantum mechanical formulation of the dipolar
susceptibility tensors they further pointed out that even a few
percent dispersion can seriously damage the validity of Kleinman
symmetry. It was further demonstrated by Franken and Ward that for
the case of quartz with the fundamental frequency at $694.3nm$ of
the Ruby laser, even though the dispersion is only about $2\%$,
the broken of the Kleinman symmetry can be as significant as
$30\%$ by considering the dipolar contribution
only.\cite{Franken1963RMP35p23} This undoubtedly suggested a
general breakdown of the Kleinman symmetry for general materials
and molecules. This is also why Franken and
Ward,\cite{Franken1963RMP35p23} as well as Ron
Shen,\cite{ShenNonlinearOpticsBook} called it as Kleinman's
conjecture instead of Kleinman
Symmetry.\cite{BoydNonlinearOpticsBook} Therefore, in the study of
real molecular systems, Kleinman symmetry should generally fail
under the dipole
approximation.\cite{Victor2001JOSAB18p1858,Simpson2004CPL390p8}
Recently, using the Kramers-Kronig dispersion relations, Dailey
\textit{et al.} again discussed this general failure of Kleinman
symmetry relationship in nonlinear susceptibility tensors of
molecular systems, and they further listed numerous experimental
disagreements with the Kleinman symmetry relationship under
non-resonant conditions in the
literatures.\cite{Simpson2004CPL390p8} Thus, it is clear that the
broken of Kleinman symmetry can never be used as the criterion for
the existence of quadrupolar (bulk) contribution to SHG signal.
However, the validity of the Kleinman symmetry had been
overvalued\cite{Eisenthal1988JPC92P5074,Eisenthal1989CPL157p101,
Frey2001MP99p677,Girault1995JCSFT91p1763} due to the lack of
awareness about the original work by Franken and
Ward.\cite{Franken1963RMP35p23}

In Franken and Ward's discussion of quartz crystal, the
macroscopic symmetry of the crystal is the same as the microscopic
symmetry of the unit cell. However, for molecular system, the
macroscopic symmetry of the molecular system and the microscopic
symmetry of the molecule are usually not identical, especially for
the liquid interfaces which usually have rotationally isotropic
symmetry around the interface normal, no matter what symmetry each
individual molecule may possess. Therefore, the connection between
the microscopic and macroscopic Kleinman symmetry relationship is
not as straightforward as that for crystals. Recently, Simpson
\textit{et al.} have tried to bridge this gap by working on the
molecular and surface hyperpolarizability of oriented chromophores
of different symmetries and different electronic
transitions.\cite{Simpson2004CPL390p8,Simpson2002PRB66,Simpson2004JPCB108p3548}
These works can certainly provide good references for dealing with
complicated relationships of susceptibility tensors in SHG and SFG
studies of molecular interfaces.

For the neat air/water interface, Goh \textit{et al.} pointed out
that derivation of the expressions of biaxial molecule ($C_{2v}$
symmetry) might help interpretation of the SHG signal from the
air/water interface.\cite{Eisenthal1989CPL157p101} This was indeed
pursued briefly for the air/water
interface,\cite{GRPintoThesis1988} but it was ultimately carried
away by the observed temperature dependence phenomenon, which
seemingly indicated significant quadrupolar contributions to the
SHG signal.\cite{Eisenthal1988JPC92P5074,Eisenthal1989CPL157p101}
Besides, the major obstacle in the advance of SHG technique for
surface studies has always been the lack of microscopic
understanding of the surface nonlinear susceptibility
$\chi_{ijk}$.\cite{ShenARPC1989}

In this report, after brief discussion of the basic theory of SHG,
we shall first present the discussion on the connection between
the microscopic polarizability tensors of the water molecule and
the macroscopic susceptibility tensors. Then we shall discuss the
general considerations regarding the interface and bulk
contributions to the SHG from an interface. Finally we shall
present the treatment on the SHG measurement with the proper
formulation of the microscopic polarizability tensors of the water
molecule, which gives well description with the dipolar nature of
the SHG response from the neat air/water interface. The major
conclusion is that with explicit treatment of the molecular
symmetry, dipolar (interface) contribution alone can provide
satisfactory explanation to the observed SHG signal from the neat
air/water interface. We shall show that not only detailed
information on molecular orientation can be obtained from
polarization SHG measurement of the neat air/water interface, but
also such treatment can be generally applied to other molecular
interfaces as well. This new understanding of the air/water
interface can shed light on our understandings of the nonlinear
optical responses from other molecular interfaces.

\section{Theoretical background}

\subsection{Basic Theory}

The basic theory of SHG as a general surface analytical probe has
been well described in the
literatures.\cite{Shen1991PRA43p6778,ShenARPC1989,
ShenZhuang1999PRB59p12632,RaoYi2004JCP119p5226} However, detailed
treatment of the experimental data in different polarizations has
not been
usual.\cite{ZhangTG1990JOSAB,ShenZhuang1999PRB59p12632,RaoYi2004JCP119p5226,
Simpson2003ACA496p133,Frey2003MST14p508,Meech2000Langmuir16p2893}

Generally, the SHG Intensity $I(2\omega)$ reflected from an
interface is given
below.\cite{ShenARPC1989,ShenZhuang1999PRB59p12632,RaoYi2004JCP119p5226}

\begin{eqnarray}
 I(2\omega )&=&\frac{{32\pi ^{3} \omega ^{2} \sec^{2} \Omega }}{{\textit{c}_0 ^{3} n_{1}(\omega)n_{1}(\omega)n_{1}(2\omega) }}
 |\chi _{eff} |^{2} I_\omega ^2 \label{intensity}\\
 \chi _{eff}&=&[\mathbb{L}(2\omega):\hat{e}(2\omega )].\chi :
 [\mathbb{L}(\omega):\hat{e}(\omega )]. [\mathbb{L}(\omega):\hat{e}(\omega
 )]\nonumber\\\label{chi1}
 \end{eqnarray}

\noindent In Equation \ref{intensity}, $I_\omega$ is the incoming
laser intensity, $\textit{c}_{0}$ is the speed of the light in the
vacuum, and $\Omega $ is the incident angle from the surface
normal. In Equation \ref{chi1}, $\chi$ is the macroscopic
second-order susceptibility tensor, which has $3\times3\times3=27$
elements; $\hat{e}(2\omega )$ and $\hat{e}(\omega )$ are the unit
vector of the electric field at $2\omega$ and $\omega$;
$\mathbb{L}(2\omega)$ and $\mathbb{L}(\omega)$ are the tensorial
Fresnel factors for $2\omega$ and $\omega$, respectively.

It is important to realize that $\chi_{eff}$ contains all
molecular information of SHG measurement. Actually $\chi_{eff}$ is
the point product of the \textit{observation vector}
$[\mathbb{L}(2\omega):\hat{e}(2\omega )]$ and the second order
\textit{polarization vector}
$\hat{P}^{(2)}=\chi:[\mathbb{L}(\omega):\hat{e}(\omega
)].[\mathbb{L}(\omega):\hat{e}(\omega)]$. In the SHG experiment,
the field vectors of the incoming and out-going light beams are
controlled by the experimenter; once these field vectors are
fixed, the tensorial Fresnel factors are also fixed quantities. So
the SHG experiment measures the SHG intensity and provides the
information of the macroscopic susceptibility of the molecular
system.

In SHG experiment, there are three independent polarization
measurement, namely, s-in/p-out ($I{sp}$), $45^\circ$-in/s-out
($I_{45^{\circ}s}$) and the p-in/p-out ($I_{pp}$). Here, in the
experimental coordinate system ($x,y,z$), $z$ is the interface
normal, and we choose $xz$ plane as the incident plane.
Subsequently, $p$ polarization is defined as polarization within
the $xz$ plane, and $s$ is perpendicular to the $xz$ plane. If the
microscopic local field factors of the molecular layer are
considered, the $\chi_{eff}$ of these three polarizations should
take the following
forms.\cite{ShenZhuang1999PRB59p12632,ShenWeiXing2000PRE62p5160,ZhengDS_2005paper_2}

\begin{eqnarray}
\chi_{eff,sp}^{}&=&L_{zz}(2\omega)L^{2}_{yy}(\omega)l_{zz}(2\omega)l^{2}_{yy}(\omega)\sin\Omega\chi_{zyy}\nonumber\\
\chi_{eff,45^{\circ}s}^{}&=&L_{yy}(2\omega)L_{zz}(\omega)L_{yy}(\omega)\nonumber\\
&&\times\l_{yy}(2\omega)l_{zz}(\omega)l_{yy}(\omega)\sin\Omega\chi_{yzy}\nonumber\\
\chi_{eff,pp}^{}&=&L_{zz}(2\omega)L^{2}_{xx}(\omega)l_{zz}(2\omega)l^{2}_{xx}(\omega)\sin\Omega\cos^{2}\Omega\chi_{zxx}\nonumber\\
&&-2L_{xx}(2\omega)L_{zz}(\omega)L_{xx}(\omega)\nonumber\\
&&\times\l_{xx}(2\omega)l_{zz}(\omega)l_{xx}(\omega)\sin\Omega\cos^{2}\Omega\chi_{xzx}\nonumber\\
&&+L_{zz}(2\omega)L^{2}_{zz}(\omega)l_{zz}(2\omega)l^{2}_{zz}(\omega)\sin^{3}\Omega\chi_{zzz}\nonumber\\\label{ChiEff}
\end{eqnarray}

\noindent where $l_{xx}(\omega_i)$, $l_{yy}(\omega_i)$ and
$l_{zz}(\omega_i)$ are the microscopic local field factors at
frequency
$\omega_i$.\cite{ShenYe1983PRB28p4288,ShenZhuang1999PRB59p12632,ZhengDS_2005paper_2}
The detailed expression for the $L_{ii}$ and $l_{ii}$ factors can
be found
elsewhere.\cite{ShenYe1983PRB28p4288,ShenZhuang1999PRB59p12632,ShenWeiXing2000PRE62p5160,ZhengDS_2005paper_2}

Now the intensity ratio $I_{eff,45^{\circ}s}/I_{eff,sp}$ should be
the square of the ratio $\chi_{eff,45^{\circ}s}/\chi_{eff,sp}$.

\begin{eqnarray}
\frac{\chi_{eff,45^{\circ}s}}{\chi_{eff,sp}}
&=&\frac{L_{yy}(2\omega)L_{zz}(\omega)}{L_{zz}(2\omega)L_{yy}(\omega)}\times
\frac{l_{yy}(2\omega)l_{zz}(\omega)}{l_{zz}(2\omega)l_{yy}(\omega)}\times
\frac{\chi_{yzy}}{\chi_{zyy}}\nonumber\\\label{ratio}
\end{eqnarray}

\noindent So the difference of the $I_{eff,45^{\circ} s}$ and
$I_{eff,sp}$ can be either from the difference of the Fresnel
factors and local field factors at $\omega$ and $2\omega$, or from
the difference between the second order susceptibility tensor
$\chi_{yzy}$ and $\chi_{zyy}$.

As we know, when both $\omega$ and $2\omega$ are far from
resonance with the molecular electronic transitions, the
difference of the Fresnel factors and local field factors at
$\omega$ and $2\omega$ are essentially
negligible.\cite{Frey2001MP99p677,ShenYe1983PRB28p4288,ShenZhuang1999PRB59p12632}
Therefore, generally we have
$\frac{L_{yy}(2\omega)L_{zz}(\omega)}{L_{zz}(2\omega)L_{yy}(\omega)}\approx{1}$
and
$\frac{l_{yy}(2\omega)l_{zz}(\omega)}{l_{zz}(2\omega)l_{yy}(\omega)}\approx{1}$
for simple liquid interfaces far from
resonance.\cite{Frey2001MP99p677} Usually, the linear dispersion
of non-resonant materials in the optical frequency region is only
in the order of $1\%$ to $3\%$.\cite{CRCHandbook}

The non-zero elements of the macroscopic susceptibility tensors
are determined by the symmetry property of the interface. For an
achiral rotationally isotropic molecular layer, there are only 7
non-zero elements, i.e. $\chi_{zzz}$, $\chi _{zxx}=\chi _{zyy}$,
$\chi_{yzy}=\chi _{yyz}=\chi_{xzx}=\chi _{xxz}$. If Kleinman
symmetry holds additionally, we have $\chi_{zzz}$, $\chi
_{zxx}=\chi _{zyy}=\chi_{yzy}=\chi _{yyz}=\chi_{xzx}=\chi
_{xxz}$.\cite{ShenZhuang1999PRB59p12632,RaoYi2004JCP119p5226}
Therefore, if the ratio $\chi_{eff,45^{\circ}s}/\chi_{eff,sp}$ is
significantly different from unity for a simple liquid interface,
i.e. $\chi_{yzy}\neq \chi_{zyy}$ from Eq.\ref{ratio}, the
macroscopic Kleinman symmetry must be broken.

\subsection{Microscopic and Macroscopic Kleinman Symmetry}

Here we discuss the direct connection between the Kleinman
symmetry of the microscopic (molecular) polarizability tensor
elements $\beta_{lmn}$ and the macroscopic susceptibility tensor
elements $\chi_{ijk}$ of the molecular system. Here we only
consider the contribution from the dipolar terms.

Under the dipolar approximation, it is known that $\chi_{ijk}$ is
generally related to $\beta_{lmn}$ through Euler transformation
operation, which is the product of three rotational operations of
the molecular coordinates
system.\cite{ShenNonlinearOpticsBook,Brevetbook,GoldsteinBook}

\begin{eqnarray}
\chi_{ijk}=N\sum_{lmn=abc}\langle{R_{il}R_{jm}R_{kn}}\rangle\beta_{lmn}\label{chibeta}
\end{eqnarray}

\noindent where $N$ is the interface molecule density, the
operator $\langle{}\rangle$ denotes the orientational ensemble
average over the Euler rotation matrix transformation element
$R_{\lambda\lambda'}$ from the molecular coordinate
$\lambda'(a,b,c)$ to the laboratory coordinate $\lambda(x,y,z)$.
\cite{RaoYi2004JCP119p5226,WangHF2004CJCP17P362} The subscript
$(i,j,k)$ of the $\chi_{ijk}$ corresponds to the laboratory
coordinate $(x,y,z)$, and the subscript $(l,m,n)$ of the
$\beta_{lmn}$ corresponds to the molecular coordinate $(a,b,c)$.
This expression can also be written into the summation over
$\beta_{lmn}$ of each molecule in the molecular system which
contributes to $\chi_{ijk}$.

\begin{eqnarray}
\chi_{ijk}=\sum_{I=1}^{N}\sum_{lmn=abc}{R_{il}^{I}R_{jm}^{I}R_{kn}^{I}}\beta_{lmn}\label{chibetaSum}
\end{eqnarray}

In the most general case there are 27 second order susceptibility
tensor elements for $\chi_{ijk}$ as well as $\beta_{lmn}$. Because
in SHG the two fundamental frequencies are equivalent, the
relationship $\chi_{ijk}=\chi_{ikj}$ and $\beta_{lmn}=\beta_{lnm}$
are always valid. So in the most general case there are only 18
independent elements for $\chi_{ijk}$ and $\beta_{lmn}$,
respectively. Kleinman symmetry states that
$\chi_{ijk}=\chi_{jik}$ for the macroscopic susceptibility tensor
elements, and $\beta_{lmn}=\beta_{mln}$ for the microscopic
polarizability tensor elements. Kleinman symmetry can further
reduce the independent elements to the number of 10. Of course,
the number of non-zero tensor elements is much smaller than this
number.

Because the microscopic Kleinman symmetry relationship requires
$\beta_{lmn}=\beta_{mln}$, and because the product of the matrix
elements always satisfy
$R_{il}^{I}R_{jm}^{I}R_{kn}^{I}=R_{jm}^{I}R_{il}^{I}R_{kn}^{I}$,
we immediately have

\begin{eqnarray}
\chi_{ijk}&=&\sum_{I=1}^{N}\sum_{lmn=abc}{R_{il}^{I}R_{jm}^{I}R_{kn}^{I}}\beta_{lmn}\nonumber\\
&=&\sum_{I=1}^{N}\sum_{lmn=abc}{R_{jm}^{I}R_{il}^{I}R_{kn}^{I}}\beta_{mln}=\chi_{jik}\nonumber\\\label{Kleinman}
\end{eqnarray}

This simple transformation proves that the Kleinman symmetry is
invariant under arbitrary rotational transformation. Because the
rotational transformation is linearly reversible, if the
macroscopic Kleinman symmetry holds, the microscopic Kleinman
symmetry should also hold, as long as all the molecules in the
system can be treated identical. This relationship conceptually
simplifies many things when we are trying to compare experimental
results with molecular properties in SHG, as well as SFG studies.

Microscopic Kleinman symmetry is valid for uniaxial rod-like
molecule, because such molecule has only one non-zero molecular
polarizability tensor element $\beta_{ccc}$. Therefore, any
macroscopic system consists of uniaxial molecule shall observe
Kleinman symmetry. It was just because the `ubiquitous' usage of
uniaxial molecule model in description of nonlinear chromophores,
Kleinman symmetry has been generally held true in the past forty
years, pointed out precisely by Dailey \textit{et
al.},\cite{Simpson2004CPL390p8} despite the seminal discussion of
Franken and Ward in 1963 on the general failure of Kleinman
symmetry even for quartz crystal.\cite{Franken1963RMP35p23}

\subsection{SHG from Liquid Interface of Water: the Molecule with $C_{2v}$ Symmetry}

Here we present the detailed treatment of the SHG from neat
air/water interface considering only the dipolar contribution.

The water molecular belongs to $C_{2v}$ symmetry. Therefore, there
are seven non-zero microscopic polarizability tensor elements in
SHG, namely, $\beta_{ccc}$, $\beta_{caa}$,
$\beta_{aca}=\beta_{aac}$, $\beta_{cbb}$,
$\beta_{bcb}=\beta_{bbc}$.\cite{ShenARPC1989,BuckinghamJOSAB1998,Brevetbook}
Here the molecular coordinates are defined to have $c$ axis along
the $C_{2}$ axis, and to have $ac$ plane as the molecular plane.
For the rotationally isotropic neat air/water interface, there are
7 non-zero elements, i.e. $\chi_{zzz}$, $\chi _{zxx}=\chi _{zyy}$,
$\chi_{yzy}=\chi _{yyz}=\chi_{xzx}=\chi _{xxz}$. Because the
twisted angle $\psi$ and the azimuthal angle $\phi$ can be
considered isotropic, using Eq.\ref{chibeta} and average over
Euler angle $\psi$ and $\phi$, we have

\begin{eqnarray}
\chi_{zzz}&=&\frac{N_s}{2}[(\beta_{caa}+\beta_{cbb}+2\beta_{aca}+2\beta_{bcb})\langle{\cos\theta}\rangle\nonumber\\
&&+(2\beta_{ccc}-\beta_{caa}-\beta_{cbb}-2\beta_{aca}-2\beta_{bcb})\langle{\cos^3\theta}\rangle]\nonumber\\
\chi_{zxx}&=&\frac{N_s}{4}[(2\beta_{ccc}+\beta_{caa}+\beta_{cbb}-2\beta_{aca}-2\beta_{bcb})\langle{\cos\theta}\rangle\nonumber\\
&&-(2\beta_{ccc}-\beta_{caa}-\beta_{cbb}-2\beta_{aca}-2\beta_{bcb})\langle{\cos^3\theta}\rangle]\nonumber\\
\chi_{xzx}&=&\frac{N_s}{4}[(2\beta_{ccc}-\beta_{caa}-\beta_{cbb})\langle{\cos\theta}\rangle\nonumber\\
&&-(2\beta_{ccc}-\beta_{caa}-\beta_{cbb}-2\beta_{aca}-2\beta_{bcb})\langle{\cos^3\theta}\rangle]\nonumber\\
\label{Chi1}
\end{eqnarray}

\noindent Because $\beta_{caa}$ and $\beta_{cbb}$ always go
together in these expressions, if $\beta_{ccc}\neq 0$, we define
$r=\frac{\beta_{caa}+\beta_{cbb}}{2\beta_{ccc}}$; and similarly we
define $s=\frac{\beta_{aca}+\beta_{bcb}}{2\beta_{ccc}}$. Then we
have

\begin{eqnarray}
\chi_{zzz}&=&{N_s}\beta_{ccc}[(r+2s)\langle{\cos\theta}\rangle+(1-r-2s)\langle{\cos^3\theta}\rangle]\nonumber\\
\chi_{zxx}&=&\frac{N_s}{2}\beta_{ccc}[(1+r-2s)\langle{\cos\theta}\rangle-(1-r-2s)\langle{\cos^3\theta}\rangle)]\nonumber\\
\chi_{xzx}&=&\frac{N_s}{2}\beta_{ccc}[(1-r)\langle{\cos\theta}\rangle-(1-r-2s)\langle{\cos^3\theta}\rangle]\nonumber\\
\label{ChiRS}
\end{eqnarray}

\noindent And the $\chi_{zxx}^{}/\chi_{xzx}^{}$ and
$\chi_{zzz}^{}/\chi_{xzx}^{}$ ratios are

\begin{eqnarray}
\frac{\chi_{zxx}^{}}{\chi_{xzx}^{}}&=&\frac{(1+r-2s)D-(1-r-2s)}{(1-r)D-(1-r-2s)}\label{Ratio1}\\
\frac{\chi_{zzz}^{}}{\chi_{xzx}^{}}&=&\frac{2(r+2s)D+2(1-r-2s)}{(1-r)D-(1-r-2s)}\label{Ratio2}
\end{eqnarray}

\noindent in which
$D=\langle{\cos\theta}\rangle/\langle{\cos^3\theta}\rangle$ is the
orientational parameter of the water molecule at the neat
air/water interface.\cite{RaoYi2004JCP119p5226} It is clear from
Eq.\ref{Ratio1} that if $r=s$, then $\chi_{zxx}^{}=\chi_{xzx}^{}$;
and \textit{vice versa}. From the definition, $r=s$ is equivalent
to $\beta_{caa}+\beta_{cbb}=\beta_{aca}+\beta_{bcb}$, which is the
necessary condition for Kleinman symmetry.

When $r\neq s$, it is clear from Eq.\ref{Ratio1} that the ratio
$\chi_{zxx}/\chi_{xzx}\neq 1$ unless $D=0$. However, $D=0$ is
physically impossible, therefore, a trivial
solution.\cite{RaoYi2004JCP119p5226} Thus, $\chi_{zxx}/\chi_{xzx}$
value not only depends on r and s, but also on the value of the
orientational parameter $D$. Only when $D=1$, we have
$\chi_{zxx}/\chi_{xzx}=r/s$, which means that the macroscopic and
microscopic tensor ratios are equal. Because the two ratios of
$\chi_{zxx}^{}/\chi_{xzx}^{}$ and $\chi_{zzz}^{}/\chi_{xzx}^{}$
can be obtained from experiment by using Eq.\ref{ChiEff}, and if
we know the value for $r/s$, $D$ value can be determined by
solving the Eq.\ref{Ratio1} and Eq.\ref{Ratio2}.

Here with the special case that when $\beta_{ccc}=0$, only the
ratio
$R=r/s=\frac{\beta_{caa}+\beta_{cbb}}{\beta_{aca}+\beta_{bcb}}$ is
physically meaningful, and Eq.\ref{Ratio1} and Eq.\ref{Ratio2}
become into the followings.

\begin{eqnarray}
\frac{\chi_{zxx}^{}}{\chi_{xzx}^{}}&=&\frac{(R-2)D+(R+2)}{-R\ast D+(R+2)}\label{Ratio10}\\
\frac{\chi_{zzz}^{}}{\chi_{xzx}^{}}&=&\frac{2(R+2)D-2(R+2)}{-R\ast
D+(R+2)}\label{Ratio20}
\end{eqnarray}

It has been known that from quantum mechanical treatment, $C_{2v}$
molecule must have a low-lying excited electronic state with a
transition dipole moment perpendicular or close to perpendicular
to the molecular dipole, i.e. along $a$ or $b$
axis.\cite{Victor2001JOSAB18p1858} This low-lying state belongs to
the $B$-type irreducible representation, and the dominant
molecular polarizability tensors for the $B$-type transition can
only be either $\beta_{aac}=\beta_{aca}$ ($B_{1}$) or
$\beta_{bbc}=\beta_{bcb}$ ($B_{2}$) in
resonance.\cite{Simpson2004JPCB108p3548} For water molecule, the
first two electronic excited states belongs to $B_{1}$ and $B_{2}$
symmetry, and the transitions are around 140nm-190nm (diffusive)
and 124nm (strong), respectively.\cite{littlebook} According to
Dailey \textit{et al.},\cite{Simpson2004CPL390p8} the
Kramers-Kronig dispersion relations dictate that even far from
resonance, the largest tensor elements for water molecule have to
be $\beta_{aac}=\beta_{aca}$ or $\beta_{bbc}=\beta_{bcb}$, and the
Kleinman symmetry has to be broken. Importantly, because the first
two transitions are along $a$ or $b$ directions, $\beta_{ccc}$ is
expected to be very small or negligible, comparing to
$\beta_{aac}=\beta_{aca}$, $\beta_{bbc}=\beta_{bcb}$,
$\beta_{caa}$ and $\beta_{cbb}$ terms, according to the quantum
mechanical sum-over-states (SOS) treatment of the molecular
hyperpolarizabilities.\cite{WardRMP1965,Franken1963RMP35p23,Simpson2004JPCB108p3548}
Therefore, the broken of macroscopic Kleinman symmetry for the
neat air/water interface reported previously is unavoidable, and
can be fully formulated using only dipolar contribution as
described in this section.

If $\beta_{ccc}$ term can be very small or negligible as discussed
above, Eq.\ref{Ratio10} and Eq.\ref{Ratio20} can be used to
extract unique $R$ and $D$ values.

\subsection{Beyond the Dipolar Contribution}

Here we discuss the implication of the above treatment on the
dipolar contribution to the total SHG signal.

Detailed treatment on contributions to the total SHG signal from
the dipole, quadrupole and field gradient across the interface
layer has been well established in the
literatures.\cite{HeinzBook,Girault1995JCSFT91p1763,ShenGuyotSionnestPRB1986,
ShenGuyotSionnestPRB1987,ShenGuyotSionnestPRB1988} The dipolar
contribution is usually referred as the interface or local
contribution, and the other two contributions are generally
referred as the bulk or nonlocal
contributions.\cite{ShenGuyotSionnestPRB1987} When the upper phase
is air, we have,

\begin{eqnarray}
\chi_{zzz}^{total}&=&\chi_{s,zzz}^{}-(\gamma+\chi_{Q,zzzz}^{})\frac{\epsilon{'}
(2\omega)\epsilon'^{2}(\omega)}{\epsilon(2\omega)\epsilon^2(\omega)}\nonumber\\
\chi_{zxx}^{total}&=&\chi_{s,zxx}^{}-(\gamma+\chi_{Q,zzxx}^{})\frac{\epsilon{'}
(2\omega)}{\epsilon(2\omega)}\nonumber\\
\chi_{xzx}^{total}&=&\chi_{s,xzx}^{}-\chi_{Q,zxzx}^{}\frac{\epsilon{'}(\omega)}
{\epsilon(\omega)}\label{ChiAll}
\end{eqnarray}

\noindent where $\chi_{ijk}^{total}$ is the total susceptibility
tensor element, which includes both the interface and bulk
contributions. $\chi_{s,ijk}$ is the dipolar susceptibility
tensor, denoted as $\chi_{ijk}$ in previous sections. $\gamma$
represents the contribution from the field gradient across the
interface. $\epsilon'(\omega)$, $\epsilon'(2\omega)$ are the
effective optical dielectric constants of the interface layer, and
$\epsilon_{}(\omega)$, $\epsilon_{}(2\omega)$, are the optical
dielectric constants of the bulk liquid. $\chi_{Q,ijkl}^{}$ is the
quadrupolar susceptibility tensor components of the bulk medium.

As we have known, the ratios between the three
$\chi_{ijk}^{total}$ in Eq.\ref{ChiAll} can be determined directly
from the SHG measurements. However, in principle the relative
contributions to $\chi_{ijk}^{total}$ from the local and nonlocal
terms can not be explicitly
estimated.\cite{ShenGuyotSionnestPRB1988,HeinzPRA1990,HeinzBook}
From Eq.\ref{ChiAll} Here we argue that if the ratios of the
$\chi_{ijk}^{total}$ terms can be quantitatively explained with
the dipolar terms, i.e. $\chi_{s,ijk}$ terms, the contributions
from the bulk terms can be very small or negligible. At least,
based on the discussions we have so far, all the previous
arguments for them to be comparable to the dipolar terms can not
be substantiated.

\section{Experimental}

A broadband tunable mode-locked femtosecond Ti:Sapphire laser
system (Tsunami, Spectra-Physics) is used for reflected-geometry
SHG measurement. Its high-repetition rate (82MHz) and short
pulse-width (80fs) make it suitable for detection of weak second
harmonic signals. Its long term power and pulse-width stability
also make it easy for quantitative analysis of the SHG
data.\cite{Hongfei2004JPCAHRS} The optical set-up is typical for
SHG experiment.\cite{RaoYi2004JCP119p5226} The 800nm fundamental
laser beam is focused on the air/water interface at the incident
angle of $\Omega=70^\circ$, and the SHG signal at 400nm is
detected with a high gain photo-multiplier tube (R585, Hamamatsu)
and a photon counter (SR400, Stanford). Typically the dark noise
level is less than 3 counts/second. The laser power is typically
500mw. The efficiency of the detection system for $p$ polarization
is 1.31 times of that for $s$ polarization. A teflon beaker
(diameter around 10cm) is used to contain the pure water liquid
(Millipore 18.2M$\Omega \cdot $cm). Optical polarization
controlling, SHG data acquisition are programmed and controlled
with a PC. It takes about 300 seconds to collect the polarization
curve. The room temperature is controlled at
$22.0\pm1.0^{\circ}$C, and the humidity of the room is controlled
around $40\%$.

\section{Results and Discussions}

\subsection{Calculation of SHG Intensities}

Here we present the analysis of the polarization SHG data with the
dipolar formulations in Section II.

\begin{figure}[h]
\begin{center}
\includegraphics[width=7.0cm]{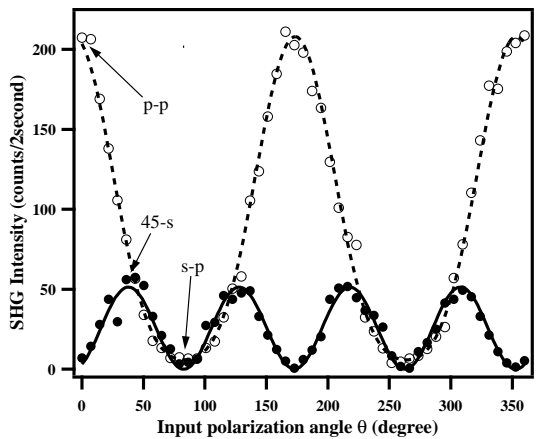}
\caption{$360^{\circ}$ polarization SHG measurement of the neat
air/water interface at two fixed detection polarizations (open
circle for $p$ detection; closed circle for $s$ detection). Each
point was averaged for three times. The lines are fitting
curves.}\label{fig1}
\end{center}
\end{figure}

The SHG signal in both $p$ and $s$ polarizations are measured with
automatic scanning of the input polarization in the full
$360^{\circ}$, as presented in Fig.\ref{fig1}. Because the SHG
signal from a neat air/liquid interface is usually very small,
full range scanning of the input polarization can help improving
the accuracy of the SHG intensities measurement in the desired
$sp$, $45^{\circ} s$ and $pp$ polarizations. The SHG signal is
especially small for $sp$ polarization of air/water interface.
With the symmetrical polarization curves as shown in
Fig.\ref{fig1}, such small signal can be determined with much
better accuracy than previously reported results, where input
polarization were only varied between $0^{\circ}$ to
$90^{\circ}$.\cite{Frey2001MP99p677,Girault1995JCSFT91p1763} The
polarization data with full $360^{\circ}$ scan range also provides
another advantage. It is that the offset of the polarization angle
in the SHG experiment, which can cause significant error on the
very small SHG signal in $sp$ polarization, can be accurately
determined.

The data in Fig.\ref{fig1} should follow the following
functions.\cite{ZhangTG1990JOSAB}

\begin{eqnarray}
I_{p}&=&[A\cos^{2}\theta+B\sin^{2}\theta]^2\nonumber\\
I_{s}&=&[C\sin(2\theta)]^{2}
\end{eqnarray}

\noindent Fitting results for the $p$ detection curve gave
$A=14.42\pm0.07$, $B=2.42\pm0.18$, the offset of $\theta$ is
$\theta_{0}=6.6^{\circ}\pm0.4^{\circ}$; and the $s$ detection
curve gave $C=7.17\pm0.08$ and
$\theta_{0}=7.2^{\circ}\pm0.5^{\circ}$. Such fitting results gave
the following values, $I_{pp}=208\pm2$, $I_{sp}=5.9\pm0.9$ and
$I_{45^{\circ}s}=51.4\pm1.1$. With the $s$ and $p$ detection ratio
of 1.31, and after corrections with the Fresnel factors and the
local field factors (which is essentially no influence), we obtain
$\chi_{zxx}/\chi_{xzx}=0.30\pm0.03$ and
$\chi_{zzz}/\chi_{xzx}=2.56\pm0.06$. These ratios are consistent
with the results reported in the literatures (Table
\ref{tabratios}),and the values obtained here should be with
better accuracy, as we compared the quality of SHG data with those
in the literatures, especially for the very small value of
$I_{sp}$.\cite{Eisenthal1988JPC92P5074,Eisenthal1989CPL157p101,
Frey2001MP99p677,Girault1995JCSFT91p1763} Nevertheless, because
$I_{sp}$ is very small, its value might have been overestimated in
the past literatures as discussed above.

It is to be noted that in calculating $\chi_{zxx}/\chi_{xzx}$ and
$\chi_{zzz}/\chi_{xzx}$ ratios, the assumption that $\chi_{zxx}$,
$\chi_{xzx}$ and $\chi_{zzz}$ all have the same signs was used as
in the previous
literatures.\cite{Eisenthal1988JPC92P5074,Eisenthal1989CPL157p101,
Frey2001MP99p677,Girault1995JCSFT91p1763} Goh \textit{et al.}
studied these relative
phases,\cite{Eisenthal1988JPC92P5074,GRPintoThesis1988} and our
results are consistent with their report on this issue.

\begin{table}[!t]
\caption{The $\chi_{zxx}/\chi_{xzx}$ and $\chi_{zzz}/\chi_{xzx}$
ratios from different research groups. The calculated $R$, $D$ and
$\theta$ values according to Eq.\ref{Ratio10} and Eq.\ref{Ratio20}
are also listed.}
\begin{center}
\begin{tabular}{lcccccccccccccc}
\hline & Eisenthal\cite{Eisenthal1988JPC92P5074} & Frey\cite{Frey2001MP99p677} & Girault\cite{Girault1995JCSFT91p1763} & This work\\
\hline
$\frac{\chi_{zxx}}{\chi_{xzx}}$ & 0.46 & 0.48$\pm$0.02($?$) & 0.41 & 0.30$\pm$0.03 \\
$\frac{\chi_{zzz}}{\chi_{xzx}}$ & 1.20 & 3.6$\pm$0.5 & 2.56 & 2.56$\pm$0.06 \\
R & 0.66 & 0.81$\pm$0.03 & 0.74 & 0.69$\pm$0.02 \\
D & 1.39 & 1.84$\pm$0.07 & 1.69 & 1.72$\pm$0.02 \\
$\theta$ & 32.0$^{\circ}$ & 42.5$^{\circ}\pm$1.2$^{\circ}$ & 39.8$^{\circ}$ & 40.3$^{\circ}\pm$0.4$^{\circ}$\\
\hline\label{tabratios}
\end{tabular}
\end{center}
\end{table}

\subsection{Water Molecular Hyperpolarization Tensors}

Here we try to estimate the water molecular orientation at the
neat air/water interface from the SHG data.

In Eq.\ref{Ratio1} and Eq.\ref{Ratio2}, there are three unknown
parameters, i.e. $r$, $s$ and $D$. In order to get the value of
$D$, value of either $r$, $s$ or $r/s$ needs to be known \textit{a
priori}. Otherwise, $D$ value can only be estimated for possible
$r/s$ values.

As discussed in Section II.C, $\beta_{ccc}$ term can be very small
or negligible among the 7 nonzero terms of the $C_{2v}$ water
molecule. Now we plug in $\chi_{zxx}/\chi_{xzx}=0.30\pm0.03$ and
$\chi_{zzz}/\chi_{xzx}=2.56\pm0.06$ in Table \ref{tabratios} into
Eq.\ref{Ratio10} and Eq.\ref{Ratio20}, and we obtain
$R=0.69\pm0.02$ and $D=1.72\pm0.02$ by solving these two
equations. If a $\delta$-distribution function of $\theta$ is
assumed, we have
$\theta=\arccos[(1/D)^{1/2}]=40.3^{\circ}\pm0.4^{\circ}$. The
calculation results with the previously reported
$\chi_{zxx}/\chi_{xzx}$ and $\chi_{zzz}/\chi_{xzx}$ values are
also listed in Table \ref{tabratios}. The $R$ and $D$ values
obtained from previously reported data are all quantitatively
consistent with our results.

The result $R<1$ confirms our expectation that
$\beta_{aac}=\beta_{aca}$ or $\beta_{bbc}=\beta_{bcb}$ are
dominant terms, and the Kleinman symmetry is broken. The fact,
that $\chi_{zxx}/\chi_{xzx}$ is significantly smaller than $R$,
indicated that the degree of broken of the Kleinman symmetry is
usually not the same for the microscopic polarizability tensors
and the macroscopic susceptibility tensors. From Franken and
Ward,\cite{Franken1963RMP35p23} $R$ value for a certain
fundamental frequency ($\omega$) smaller than the resonant
frequency can be estimated with the known low-lying electronic
transition frequency ($\omega_{eg}$). Franken and Ward
defined,\cite{Franken1963RMP35p23}

\begin{eqnarray}
&&|\chi_{ijk}^{(2)}-\chi_{jik}^{(2)}|\approx\bar{\epsilon}\chi_{ijk}\nonumber\\
&&\bar{\epsilon}\approx\frac{2\omega^{2}+\omega_{eg}^{2}}{\omega_{eg}^{2}-4\omega^2}-1\label{Franken}
\end{eqnarray}

\noindent Clearly, $R=1/(1+\bar{\epsilon})$. In our calculation,
with laser wavelength as 800nm, and water transition wavelength as
124nm, we obtained $R=0.86$; with laser wavelength as 800nm, and
water transition wavelength as 190nm, $R=0.70$. These $R$ values
well agree with our experimental value.

Theoretical calculations have generated scattered values for
ratios of hyperpolarizability tensors for the water molecule, when
the possibility for broken of Kleinman symmetry was
considered.\cite{Computation_JCP98p7159_1993,Computation_JCP109p5576_1998,
Computation_JCP113p1813_2000,Computation_JCP119p10519_2003,MikkelsenJPCA2004WaterCompu}
In many of other calculations, Kleinman symmetry was used as a
constraining condition. $\beta_{ccc}$ value in all these
calculations are usually comparable or even larger than other
tensor elements. However, results vary significantly with
calculation models used.\cite{MikkelsenJPCA2004WaterCompu} So far,
experimental values was scarce for comparison with these
calculations.\cite{MikkelsenJPCA2004WaterCompu} No experimental
value for ratio of hyperpolarizability tensors has been available.
The experimental value of $R$ we obtained here can be used for
such comparison with calculated results.

In our calculation, if $\beta_{ccc}$ can not be neglected, $r$,
$s$ and $D$ can not be uniquely solved with Eq.\ref{Ratio1} and
Eq.\ref{Ratio2}. If we let $r=0.5$, which means
$2\beta_{ccc}=\beta_{caa}+\beta_{cbb}$, then we get
$s=0.95\pm0.04$, i.e. $R=0.53\pm0.02$, and $D=1.81\pm0.02$, i.e.
$\theta=42.0^{\circ}\pm0.3^{\circ}$, from
$\chi_{zxx}/\chi_{xzx}=0.30\pm0.03$ and
$\chi_{zzz}/\chi_{xzx}=2.56\pm0.06$. If we allow $\beta_{ccc}$ to
be even larger, for example, we let $r=0.25$, then we get
$s=0.58\pm0.03$, i.e. $R=0.43\pm0.02$, and $D=2.08\pm0.03$, i.e.
$\theta=46.1^{\circ}\pm0.4^{\circ}$. These values indicated that
even though $r$, $s$ and $D$ values can not be uniquely determined
with experimental values of $\chi_{zxx}/\chi_{xzx}$ and
$\chi_{zzz}/\chi_{xzx}$, the $R$ and $D$ values are not very
sensitive to the nonzero $\beta_{ccc}$ values used. Especially,
the calculated $\theta$ value changed only less than 6 degrees.

Here we discussed how to extract polarizability tensor ratio $R$,
or $r$ and $s$, from the SHG experimental data considering of
dipolar (local or interface) contribution only. The values we
obtained are self-consistent, and can provide a good microscopic
description of the SHG polarization signal from the neat air/water
interface. These analyses indicate that quadrupole (nonlocal or
bulk) contributions to SHG signal from air/water interface can be
very small or negligible, even though we can not say that such
contributions has to be negligible. Most importantly, these
analyses demonstrate that the previous rationales and
evidences\cite{Eisenthal1988JPC92P5074,Eisenthal1989CPL157p101,
Frey2001MP99p677,Girault1995JCSFT91p1763} can no longer be used to
draw conclusion for significant contributions of the bulk terms to
the SHG signal from the neat air/water interface.

\subsection{Water Molecule Orientation at Neat Air/Water Interface}

Here we discuss the orientation of water molecule at the neat
air/water interface in the light of the SHG analysis above.

As we have known, there are at least two kinds of water molecules
at the neat air/water interface from the SFG-VS experimental
studies.\cite{ShenDuQ1993PRL70p2313,RichmondARPC2001,Richmond:cr102:2693,
watershortpaper,waterlongpaper} Namely, one kind (straddle-type)
is with one free O-H bond sticking out of the liquid interface and
with its dipole vector (from positive to negative) point almost
parallel to the air/water interface; and the other kind
(non-staddle-type) is with both O-H bond hydrogen-bonded and the
dipole vector point away from the liquid
phase.\cite{Richmond-jpca2000,waterlongpaper} For straddle-type
interfacial water molecule, both the free O-H bond and the
hydrogen-bonded O-H bond can only be treated as $C_{\infty v}$
symmetry
vibrationally.\cite{ShenDuQ1993PRL70p2313,watershortpaper,waterlongpaper}
While the non-staddle-type water molecule belongs to the $C_{2v}$
symmetry.\cite{watershortpaper,waterlongpaper} The vibrational
spectra of the stretching modes of both types of interfacial water
molecules have been identified and the molecular orientation has
been
studied.\cite{ShenDuQ1993PRL70p2313,RichmondARPC2001,Richmond:cr102:2693,
watershortpaper,waterlongpaper}

However, it is that the electronic transitions are responsible for
the SHG responses. Electronically, it is expected that the
straddle-type water molecule should still possesses most signature
of $C_{2v}$ symmetry, because the hydrogen bonding to one of the
O-H bond perturbs much less significantly to the electronic
transition, which has much higher transition energy and much
broader spectra, than to the vibrational transition. Therefore, it
is easy to conclude that the straddle-type water molecule should
give very small SHG signal, with the dipole vector almost parallel
to the interface. Du \textit{et al.} have shown that the
straddle-type water molecule consists about 20$\%$ of the
interfacial water molecules at the neat air/water
interface.\cite{ShenDuQ1993PRL70p2313} Thus, it is easy to see
that the SHG signal comes almost all from the non-straddle-type
water molecule. So, the $D$ values determined in Section IV.B
above provide orientational information only for the
non-straddle-type water molecule at the neat air/water interface.

The orientational parameter $D$ obtained above indicated that the
dipole vector of the non-straddle-type interfacial water molecule
oriented around an apparent orientational angle of $40^{\circ}$
from the interface normal. However, this angle was calculated
assuming a $\delta$ distribution function of the tilt angle
$\theta$. As Simpson and Rowlen have shown, such an orientational
angle is close to the `Magic Angle'
$\theta=39.2^{\circ}$.\cite{SimpsonJACS1999} Therefore, it may
come from a relatively broad orientational distribution. Indeed,
because there are different kinds of non-straddle-type water
molecule at the interface region, as clearly indicated by the
broad vibrational spectra between $3100cm^{-1}$ and $3500cm^{-1}$
in SFG-VS
studies,\cite{ShenDuQ1993PRL70p2313,RichmondARPC2001,Richmond:cr102:2693,
watershortpaper,waterlongpaper} a relatively broad orientational
distribution for the non-straddle-type hydrogen-bonded water
molecule is generally expected. However, the distribution width
should not be extremely broad. Otherwise, SHG or SFG-VS signal
from the air/water interface should have been too weak to be
detectable. On the other hand, recent detailed SFG-VS analysis has
demonstrated that the orientational distribution of straddle-type
water molecule appears to be as narrower as less than
$15^{\circ}$.\cite{watershortpaper,waterlongpaper}

On the other hand, broad orientational distribution has been
predicted for both the straddle-type and non-straddle-type
interfacial water molecule from Molecular Dynamics (MD) and Monte
Carlo (MC) simulation
studies.\cite{Wilson1987JPCWaterSimulation,RiceJCP1991,Yang_1991JPCM3pF109,
BenjaminPRL1994,Besseling_1994JPC98p11610,Garrett_1996JPC100p11720,
Sokhan_1997MP92p625,Fradin_2000Nature403p871,HynesCP2000,
MooreJCP2003,Mundy-science,JaqamanJCP2004WaterOrientationOrder}
Sokhan and Tildesley discussed the apparent disagreement of the
early SHG and SFG results for the neat air/water interface
structure.\cite{Sokhan_1997MP92p625} Goh \textit{et al.}'s early
SHG studies indicated that water hydrogens are pointing towards
the liquid side.\cite{Eisenthal1988JPC92P5074} While Du \textit{et
al.}'s SFG-VS study identified the staddle-type water molecule at
the interface. From their simulation results, Sokhan and Tildesley
believed that SHG and SFG-VS techniques are sensitive to different
parts of the interface, respectively. Simulation results often
found that the plane of water molecule is aligned approximately
parallel to the interface on the liquid side and perpendicular to
the interface on the gas side, and the dominant contribution to
the interface susceptibility is from the liquid
side.\cite{Wilson1987JPCWaterSimulation,Yang_1991JPCM3pF109,
BenjaminPRL1994,Besseling_1994JPC98p11610,Garrett_1996JPC100p11720,
Sokhan_1997MP92p625,Fradin_2000Nature403p871,HynesCP2000,
MooreJCP2003,Mundy-science,JaqamanJCP2004WaterOrientationOrder}
Unfortunately, such conclusions are drastically in disagreement
with known SHG and SFG-VS experimental observations of the neat
air/water interface, because when the water molecule has its
molecular plane lying parallel to the interface, it is impossible
to detect any SHG or SFG-VS signal in any polarization and
experimental configuration. Detailed analysis of the SHG and
SFG-VS experimental data has been significantly improved and
systematically demonstrated in recent
years.\cite{ShenZhuang1999PRB59p12632,RaoYi2004JCP119p5226,Lurong1,
Lurong2,Lurong3,Shen2005PRLWaterQuartz,ChenhuaJPCBacetone,
ChenhuaJPCBmethanol,GanweiCPLNull,ChenhuaCPLacetone,WangHF2004CJCP17P362,
HongfeiIRPCreview,watershortpaper,waterlongpaper} Even though
simulation results have provided a great deal of detailed
understandings of the liquid interfaces, these disagreements
between experimental and theoretical results suggest that the
physical picture provided by the simulations may need to be
systematically reexamined as discussed below.

As we have demonstrated above, SHG probes mainly the
non-straddle-type of interfacial water molecule, which has both
O-H bond hydrogen-bonded. On the other hand, SFG-VS probes the
stretching vibrational spectra of both the straddle-type and the
non-straddle-type interfacial
molecules.\cite{ShenDuQ1993PRL70p2313,RichmondARPC2001,Richmond:cr102:2693,
watershortpaper,waterlongpaper} Recent detailed analysis of the
SFG-VS spectra in different polarizations and experimental
configurations have helped to clarify the spectral assignments,
symmetry determination of the interfacial water species, as well
as understanding more on the orientation and motion of water
molecules at the neat air/water
interface.\cite{watershortpaper,waterlongpaper} It has been known
that most of these non-straddle-type water molecules have either 4
or 3 hydrogen-bonds on
average.\cite{Richmond-jpca2000,RichmondARPC2001,Richmond:cr102:2693,
watershortpaper,waterlongpaper} Therefore, these non-straddle-type
water molecules have to reside on the liquid side instead of the
gas side as suggested by the simulation results. As probed by SHG
and SFG-VS, their molecular plane can not be close to parallel to
the interface. Of course, it is clear that the straddle-type water
molecule is at the outermost edge of the liquid boundary, and its
molecular plane is certainly close to perpendicular to the liquid
interface. However, to call this layer as the gas side of the
interface may not be appropriate, because the air/water interface
is extremely sharp ($3.2\AA$) from X-ray
reflectivity\cite{PershanPRL1985,PershanPRA1988} and theoretical
simulations.\cite{RiceJCP1985,RiceJCP1991} As the experimental
analyses have consistently given relatively narrower orientational
distribution of the liquid interfacial molecules, simulation
results have always accounted for a much broader orientational
distribution.\cite{ShenMirandaJPCBReview,Lurong1,ChenhuaJPCBacetone,
ChenhuaJPCBmethanol,GanweiCPLNull,ChenhuaCPLacetone,ShenLinJCPAcetone2001,
HorvaiJPCB2005Acetone,JaqamanJCP2004WaterOrientationOrder} Such
issue has to be answered in the future studies.

Here we make one final comment on the orientational energy of the
air/water interface. Using the temperature effect of SHG signal
they measured, Goh \textit{et al.} concluded that it is about
$\frac{1}{2}kT$, which suggested a quite disordered air/water
interface at the room temperature.\cite{Eisenthal1989CPL157p101}
Since now the temperature effect has to be gone, the value of the
orientational energy from the same treatment of Goh \textit{et
al.} has to be much bigger than a few $kT$. This nevertheless
suggests a much better ordered air/water interface, and it is
consistent with the hydrogen-bonding nature of the molecules at
the neat air/water interface.\cite{ShenMirandaJPCBReview} The
hydrogen bond energy is typically about $5kcal/mol$, and it
implies a much well ordered interface structure.

In short conclusion, non-resonant SHG provides orientational
measurement of the non-straddle-type of interfacial water
molecule. Its dipole vector orientation is distributed around
$40^{\circ}$ from the interface normal. This picture is fully
consistent with and complementary to the SFG-VS experimental
results. Molecular simulation results may need to be reexamined
for detailed orientational structure of the neat air/water
interface.

\section{Conclusion}

As Shen had insightfully pointed out, the major obstacle in the
advance of the (SHG) technique for surface studies is the lack of
microscopic understanding of the surface nonlinear susceptibility
$\chi_{s}$.\cite{ShenARPC1989} However, the over simplifying
uniaxial molecular model has been generally used in the field. In
recent years, efforts for systematic treatment of the problem
beyond the uniaxial molecular model have been attempted by Simpson
and co-workers.\cite{Simpson2004JPCB108p3548,Simpson2002PRB66} Our
recent experiences in quantitative treatment of the SFG-VS
spectra, which always involves treatment of non-uniaxial molecular
groups, prompted us to work similarly on the treatment of SHG of
molecular interfaces.\cite{RaoYi2004JCP119p5226,Lurong1,
Lurong2,Lurong3,ChenhuaJPCBacetone,ChenhuaJPCBmethanol,
GanweiCPLNull,ChenhuaCPLacetone,WangHF2004CJCP17P362,
HongfeiIRPCreview,watershortpaper,waterlongpaper}

In this work, we re-examined the problem of the SHG reflected from
neat air/water interface with detailed treatment on the connection
between the microscopic polarizability of water molecule and
macroscopic susceptibility of the interface. Because these
treatment on microscopic polarizability is based on the molecular
symmetry properties of single water molecule, so it can be well
suitable for treating water molecule, as well as other molecules
with the same symmetry, at different molecular interfaces. After
examination on issues of the origin and validity of the Kleinman
symmetry, we have concluded that the broken of macroscopic
Kleinman symmetry of the neat air/water interface is not an
indication for quadrupole (bulk) contribution to the surface SHG
signal, as previous studies suggested and generally believed in
the field. We further demonstrated that using dipole contribution
only can fully address the broken of macroscopic Kleinman symmetry
of the air/water interface. Using properly combined molecular
polarizability tensor terms, we further simplified the treatment
and are able to obtain the value of orientational parameter $D$
from SHG data. Such procedures can be generally transferred to the
treatment of interfacial molecules belong to different molecular
symmetry.

In conclusion, the SHG signal from the neat air/water interface
can be treated fully with dipole (local or interface)
contribution. The broken of macroscopic Kleinman symmetry is not a
sufficient condition for quadrupole (nonlocal or bulk)
contributions in surface SHG. Even though in general the
quadrupole (bulk) contributions has to be considered in SHG
studies, this work suggests that they can be very small or even
negligible for the neat air/water interface, which has long been
considered the primary case favoring dominant quadrupole (bulk)
contributions. The successful treatment of the neat air/water
interface with dipole contribution terms indicates that SHG
probably is indeed a probe of the surfaces and interfaces of
isotropic fluids, as strongly argued by Andrews \textit{et
al}.\cite{AndrewsPRA1988,ShenPRA1990,AndrewsPRA1990,HeinzPRA1990,
AndrewsJModOpt1993} The treatment and approaches we presented here
can be applied and tested with other fluid interfaces.

\textbf{Acknowledgment} This work was supported by the Natural
Science Foundation of China (NSFC, No.20425309) and the Chinese
Ministry of Science and Technology (MOST, No.G1999075305). We
thank Ke-xiang Fu and Xiang-yuan Li for discussions on Section
II.B.




\begin{thebibliography}{99}
\bibitem{Shen1983PRA28p1883}T. F. Heinz, H. W. K. Tom, and Y. R. Shen, Phys. Rev. A \textbf{28}, 1883-1885 (1983).
\bibitem{Richmond1985Review}G. L. Richmond, J. M. Robinson, and V. L. Shannon, Prog. Surf. Sci. \textbf{28}, 1-70 (1988).
\bibitem{Eisenthal1988JPC92P5074}M. C. Goh, J. M. Hicks, K. Kemnitz, G. R. Pinto, K. Bhattacharyya, K. B. Eisenthal, and T. F. Heinz, J. Phys. Chem. \textbf{92}, 5074-5075 (1988).
\bibitem{Eisenthal1989CPL157p101}M. C. Goh and K. B. Eisenthal, Chem. Phys. Lett. \textbf{157}, 101-104 (1989).
\bibitem{ShenARPC1989}Y. R. Shen, Annu. Rev. Phys. Chem. \textbf{40}, 327-350 (1989).
\bibitem{EisenthalAccChemRes1993}K. B. Eisenthal, Acc. Chem. Res. \textbf{26}, 636-643 (1993).
\bibitem{Eisenthal1993CPL202p513}X. L. Zhao, S. W. Ong, and K. B. Eisenthal, Chem. Phys. Lett. \textbf{202}, 513-520 (1993).
\bibitem{Richmond1994JPC98p9688}J. C. Conboy, J. L. Daschbach, and G.L.Richmond, J. Phys. Chem. \textbf{98}, 9688-9692 (1994).
\bibitem{Girault1995JCSFT91p1763}A. A. T. Luca, P. Hebert, P. F. Brevet, and H. H. Girault, J. Chem. Soc., Faraday Trans. \textbf{91}, 1763-1768 (1995).
\bibitem{Eisenthal1996ChemRev}K. B. Eisenthal, Chem. Rev. \textbf{96}, 1343-1360 (1996).
\bibitem{Girault1997JCSFT93p3833}R. Antoine, F. Bianchi, P. F. Brevet, and H. H. Girault, J. Chem. Soc., Faraday Trans. \textbf{93}, 3833-3838 (1997).
\bibitem{Frey2001MP99p677}A. J. Fordyce, W. J. Bullock, A. J. Timson, S. Haslam, R. D. Spencer-Smith, A. Alexander, and J. G. Frey, Molec. Phys. \textbf{99}, 677-687 (2001).
\bibitem{ShenGuyotSionnestPRB1986}P. Guyot-Sionnest, W. Chen, and Y.R.Shen, Phys. Rev. B \textbf{33}, 8254-8263 (1986).
\bibitem{ShenGuyotSionnestPRB1987}P. Guyot-Sionnest and Y. R. Shen, Phys. Rev. B \textbf{35}, 4420-4426 (1987).
\bibitem{ShenGuyotSionnestPRB1988}P. Guyot-Sionnest and Y. R. Shen, Phys. Rev. B \textbf{38}, 7985-7989 (1988).
\bibitem{AndrewsPRA1988}D. L. Andrews and N. P. Blake, Phys. Rev. A \textbf{38}, 3113-3115 (1988).
\bibitem{ShenPRA1990}X. D. Zhu and Y. R. Shen, Phys. Rev. A \textbf{41}, 4549-4549 (1990).
\bibitem{AndrewsPRA1990}D. L. Andrews and N. P. Blake, Phys. Rev. A \textbf{41}, 4550-4551 (1990).
\bibitem{HeinzPRA1990}T. F. Heinz and D. P. Divincenzo, Phys. Rev. A \textbf{42}, 6249-6251 (1990).
\bibitem{AndrewsJModOpt1993}D. L. Andrews, J. Mod. Opt. \textbf{40}, 939-946 (1993).
\bibitem{HeinzBook}T. F. Heinz, \textit{Nonlinear Optical Effects at Surfaces and Interfaces},
in \textit{Nonlinear Surface Electromagnetic Phenomena}, ed. by H.
E. Ponath and G. I. Stegman (North-Holland, Armsterdam, 1991).
p353-416.
\bibitem{ShenAppliedPhysics1999}Y. R. Shen, Appl. Phys. B. \textbf{68}, 295-300 (1999).
\bibitem{ShenheldPRB2002}H. Held, A. I. Lvovsky, X. Wei, and Y. R. Shen, Phys. Rev. B \textbf{66}, 205110 (2002).
\bibitem{MoritaCPL2005SFG}A. Morita, Chem. Phys. Lett. \textbf{398},
361-366 (2004).
\bibitem{Kleinman_1962PR126p1977}D. A. Kleinman, Phys. Rev. \textbf{126}, 1977-1979 (1962).
\bibitem{ShenDuQ1993PRL70p2313}Q. Du, R. superfine, E. Freysz, and Y.R.Shen, Phys. Rev. Lett. \textbf{70}, 2313-2316 (1993).
\bibitem{Notebook}From Hong-fei Wang's notebook kept in the Eisenthal
group. Recorded between January 25, 1993 and February 10, 1993.
\bibitem{BloembergenPR1968}N. Bloembergen, R. K. Chang, S. S. Jha, and C. H. Lee, Phys. Rev. \textbf{174}, 813-822 (1968).
\bibitem{BrownPhysRev1969}F. Brown and M. Matsuoka, Phys. Rev.
\textbf{185}, 985-987 (1969).
\bibitem{ARPC1989Cited12}C. K. Chen, A. R. B. de Castro, and Y.
R. Shen, Phys Rev. Lett. \textbf{46}, 145-148 (1981)
\bibitem{ARPC1989Cited131}C. K. Chen, T. F. Heinz, D. Ricard, and Y.
R. Shen, Phys. Rev. Lett. \textbf{46}, 1010-1012 (1981)
\bibitem{ARPC1989Cited132}C. K. Chen, T. F. Heinz, D. Ricard, and Y.
R. Shen, Phys. Rev. B \textbf{27}, 1965-1979 (1983)
\bibitem{ARPC1989Cited14}T. F. Heinz, C. K. Chen, D. Ricard, and
Y. R. Shen, Phys. Rev. Lett. \textbf{48}, 478-481 (1982)
\bibitem{BoydNonlinearOpticsBook}R. W. Boyd, \textit{Nonlinear Optics} (Academic, San Diego, CA, 1992).
\bibitem{Franken1963RMP35p23}P. A. Franken and J. F. Ward, Rev. Mod. Phys. \textbf{35}, 23-39 (1963).
\bibitem{Simpson2004CPL390p8}C. A. Dailey, B. J. Burke, and G. J. Simpson, Chem. Phys. Lett. \textbf{390}, 8-13 (2004).
\bibitem{ShenNonlinearOpticsBook}Y. R. Shen, \textit{The Principles of Nonlinear Optics} (Wiley, New York, 2003).
\bibitem{Victor2001JOSAB18p1858}V. Ostroverkhov, K. D. Singer, and R. G. Petschek, J. Opt. Soc. Am. B \textbf{18}, 1858-1865 (2001).
\bibitem{Simpson2002PRB66}G. J. Simpson, J. M. Perry, and C. L. Ashmore-Good, Phys. Rev. B \textbf{66}, 165437 (2002).
\bibitem{Simpson2004JPCB108p3548}A. J. Moad and G. J. Simpson, J. Phys. Chem. B \textbf{108}, 3548-3562 (2004).
\bibitem{GRPintoThesis1988}G. R. Pinto, Ph.D. Dissertation,
\textit{Nonlinear Optics as a Probe of Structure and Dynamics at
Liquid Surfaces}, Department of Chemistry, Columbia University
(1988).
\bibitem{Shen1991PRA43p6778}M. B. Feller, W. Chen, and Y. R. Shen,
Phys. Rev. A \textbf{43}, 6778-6792 (1991).
\bibitem{ShenZhuang1999PRB59p12632}X. Zhuang, P. B. Miranda, D. Kim, and Y. R. Shen, Phys. Rev. B \textbf{59}, 12632-12640 (1999).
\bibitem{RaoYi2004JCP119p5226}Y. Rao, Y. S. Tao, and H. F. Wang, J. Chem. Phys. \textbf{119}, 5226-5236 (2003).
\bibitem{ZhangTG1990JOSAB}T. G. Zhang, C. H. Zhang, and G. K.
Wong, J. Opt. Soc. Am. B \textbf{7}, 902-907 (1990)
\bibitem{Simpson2003ACA496p133}R. M. Plocinik and G. J. Simpson, Ana. Chim. Acta \textbf{496},
133-142 (2003).
\bibitem{Frey2003MST14p508}A. J. Timson, R. D. Spencer-Smith,
A. K. Alexander, R. Greef, and J. G. Frey, Meas. Sci. Technol.
\textbf{14}, 508-515 (2003).
\bibitem{Meech2000Langmuir16p2893}S. J. Lin and S. R. Meech,
Langmuir \textbf{16}, 2893-2898 (2000).
\bibitem{ShenWeiXing2000PRE62p5160}X. Wei, S. C. Hong, X. W. Zhuang, T. Goto, and Y. R.
Shen, Phys. Rev. E \textbf{62}, 5160-5172 (2000).
\bibitem{ZhengDS_2005paper_2}D. S. Zheng and H. F. Wang, J. Chem. Phys. (to be submitted).
\bibitem{ShenYe1983PRB28p4288}P. X. Ye and Y. R. Shen, Phys. Rev. B \textbf{28}, 4288-4294
(1983).
\bibitem{CRCHandbook}D. R. Lide, \textit{CRC Handbook of Chemistry and
Physics}, (CRC Press, 81st ed., New York, 2000).
\bibitem{Brevetbook}P. F. Brevet, \textit{Surface Second Harmonic Generation} (Press Polytechniques et
Universitaires Romandes, Lausanne,1997).
\bibitem{GoldsteinBook}H. Goldstein, \textit{Classical Mechanics} (Addison-Wesley Publishing Company, Inc., 1980). pp147.
\bibitem{WangHF2004CJCP17P362}H. F. Wang, Chin. J. Chem. Phys. \textbf{17}, 362-368 (2004).
\bibitem{BuckinghamJOSAB1998}P. Fischer and  A. D. Buckingham, J. Opt. Soc. Am. B \textbf{15}, 2951-2957 (1998).
\bibitem{littlebook}(a) H. Okabe, \textit{PhotoChemistry of Small Molecules} (Wiley, New York, 1978).
(b) J. W. C. Johns, Can. J. Phys. \textbf{41}, 209-219 (1963). (c)
S. Bell, J. Mol. Spectrosc. \textbf{16}, 205-213 (1965). (d) J. A.
Horsley and W. H. Fink, J. Chem. Phys. \textbf{50}, 750-758
(1969). (e) K. J. Miller, S. R. Mielczarek, and M. Krauss, J.
Chem. Phys. \textbf{51} 26-32 (1969)
\bibitem{WardRMP1965}J. F. Ward, Rev. Mod. Phys. \textbf{37}, 1-18 (1965).
\bibitem{Hongfei2004JPCAHRS}Y. Rao, X. M. Guo, Y. S. Tao, and H. F.
Wang, J. Phys. Chem. A \textbf{108}, 7987-7982 (2004).
\bibitem{MikkelsenJPCA2004WaterCompu}K. O. Sylvester-Hvid, K. V.
Mikkelsen, P. Norman, D. Johnson, and H. $\AA$gren, J. Phys. Chem.
A \textbf{108}, 8961-8965 (2004).
\bibitem{Computation_JCP98p7159_1993}Y. Luo, H. $\AA$gren, O. Vahtras, P. J$\o$rgensen, V. Spirko, and H. Hettema, J. Chem. Phys. \textbf{98}, 7159-7164 (1993).
\bibitem{Computation_JCP109p5576_1998}K. O. Sylvester-Hvid, K. V. Mikkelsen, D. Jonsson, P. Norman, and H. $\AA$gren, J. Chem. Phys. \textbf{109}, 5576-5584 (1998).
\bibitem{Computation_JCP113p1813_2000}G. Maroulis, J. Chem. Phys. \textbf{113}, 1813-1820 (2000).
\bibitem{Computation_JCP119p10519_2003}J. Kongsted, A. Osted, K. V. Mikkelsen, and O. Christiansen, J. Chem. Phys. \textbf{119}, 10519-10535 (2003).
\bibitem{RichmondARPC2001}G. L. Richmond, Annu. Rev. Phys. Chem.
\textbf{52}, 357-389 (2001).
\bibitem{Richmond:cr102:2693}G. L. Richmond, Chem. Rev. \textbf{102}, 2693-2724 (2002).
\bibitem{watershortpaper}W. Gan, D. Wu, Z. Zhang, and H. F. Wang, Phys. Rev.
Lett. (submitted).
\bibitem{waterlongpaper}W. Gan, D. Wu, Z. Zhang, R. R. Feng, and H. F. Wang, J. Chem. Phys. (submitted).
\bibitem{Richmond-jpca2000}M. G. Brown, E. A. Raymond, H. C. Allen, L. F. Scatena, and G. L. Richmond,
J. Phys. Chem. A \textbf{104}, 10220-10226 (2000).
\bibitem{SimpsonJACS1999}G. J. Simpson, and K. L. Rowlen, J. Am. Chem. Soc. \textbf{121}, 2635-2636 (1999).
\bibitem{RiceJCP1991}R. M. Townsend and S. A. Rice, J. Chem. Phys. \textbf{94}, 2207-2218 (1991).
\bibitem{Wilson1987JPCWaterSimulation}M. A. Wilson, A. Pohorille,
and L. R. Pratt, J. Phys. Chem. \textbf{91}, 4873-4878 (1987).
\bibitem{Yang_1991JPCM3pF109}B. Yang, D. E. Sullivan, B. Tjipto-Margo, and C. G. Gray, J. Phys.
:Condens. Matter \textbf{3}, F109-F125 (1991).
\bibitem{BenjaminPRL1994}I. Benjamin, Phys. Rev. Lett. \textbf{73}, 2083-2086 (1994).
\bibitem{Besseling_1994JPC98p11610}N. A. M. Besseling and J. Lyklema, J. Phys. Chem. \textbf{98}, 11610-11622 (1994).
\bibitem{Garrett_1996JPC100p11720}R. S. Taylor, L. X. Dang, and B. C. Garrett, J. Phys. Chem \textbf{100}, 11720-11725 (1996).
\bibitem{Sokhan_1997MP92p625}V. P. Sokhan and D. J. Tildesley, Molec. Phys. \textbf{92}, 625-640 (1997).
\bibitem{Fradin_2000Nature403p871}C. Fradin, A. Braslau, D. Luzet, D. Smilgies, M. Alba, N. Boudet, K. Mecke, and J. Daillant, Nature \textbf{403}, 871-874 (2000).
\bibitem{HynesCP2000}A. Morita and J. T. Hynes, Chem. Phys. \textbf{258}, 371-390 (2000).
\bibitem{MooreJCP2003}A. Perry, H. Ahlborn, B. Space, and P. B. Moore,
J. Chem. Phys. \textbf{118}, 8411-8419 (2003).
\bibitem{Mundy-science}I-F. W. Kuo and C. J. Mundy, Science \textbf{303}, 658-660 (2004).
\bibitem{JaqamanJCP2004WaterOrientationOrder}K. Jaqaman, K.
Tuncay, and P. J. Ortoleva, J. Chem. Phys. \textit{120}, 926-938
(2004).
\bibitem{Lurong2}R. Lu, W. Gan, B. H. Wu, H. Chen, and H. F. Wang,
J. Phys. Chem. B \textbf{108}, 7297-7306 (2004).
\bibitem{Lurong3}R. Lu, W. Gan, B. H. Wu, Z. Zhang, Y. Guo, and H. F.
Wang, J. Phys. Chem. B. \textbf{109}, 14118-14129 (2005).
\bibitem{HongfeiIRPCreview}H. F. Wang, W. Gan, R. Lu, Y. Rao, and B. H. Wu, Int. Rev. Phys.
Chem. In press.
\bibitem{GanweiCPLNull}W. Gan, B. H. Wu, H. Chen, Y. Guo, and H. F. Wang, Chem. Phys.
Lett. \textbf{406}, 467-473 (2005).
\bibitem{Lurong1}(a) R. Lu, W. Gan, and H. F. Wang, Chin. Sci. Bull. \textbf{48},
2183-2187 (2003); (b) R. Lu, W. Gan, and H. F. Wang, Chin. Sci.
Bull. \textbf{49}, 899 (2004).
\bibitem{Shen2005PRLWaterQuartz}V. Ostroverkhov, G. A. Waychunas, and Y. R.
Shen, Phys. Rev. Lett. \textbf{94}, 046102 (2005).
\bibitem{ChenhuaJPCBacetone}H. Chen, W. Gan, B. H. Wu, D. Wu, Y. Guo, and H. F. Wang,
J. Phys. Chem. B, \textbf{109}, 8053-8063 (2005).
\bibitem{ChenhuaJPCBmethanol}H. Chen, W. Gan, R. Lu, Y. Guo, and H. F. Wang,
J. Phys. Chem. B, \textbf{109}, 8064-8075 (2005).
\bibitem{ChenhuaCPLacetone}H. Chen, W. Gan, B. H. Wu, D. Wu, Z. Zhang, and H. F. Wang,
Chem. Phys. Lett, \textbf{408}, 284-289 (2005).
\bibitem{PershanPRL1985}A. Braslau, M. Deutsch, P. S. Pershan, A. H.
Weiss, J. Als-Nielsen, and J. Bohr, Phys. Rev. Lett. \textbf{54},
114-117 (1985).
\bibitem{PershanPRA1988}A. Braslau, P. S. Pershan, G. Swislow, B.
M. Ocko, and J. Als-Nielsen, Phys. Rev. A \textbf{38}, 2457-2470
(1988).
\bibitem{RiceJCP1985}R. M. Townsend, J. Gryko, and S. A. Rice, J. Chem. Phys. \textbf{82}, 4391-4392 (1985).
\bibitem{ShenMirandaJPCBReview}P. B. Miranda and Y. R. Shen,
J. Phys. Chem. B \textbf{103}, 3292-3307 (1999).
\bibitem{ShenLinJCPAcetone2001}Y. L. Yeh, C. Zhang, H. Held, A. M. Mebel, X. Wei,
S. H. Lin, and Y. R. Shen, J. Chem. Phys. \textbf{114}, 1837-1843
(2001).
\bibitem{HorvaiJPCB2005Acetone}L. P\'{a}rtay and P. Jedlovszky,
and G. Horvai, J. Phys. Chem. B \textbf{109}, 12014-12019 (2005).
\end{thebibliography}
\end{document}